\documentclass[acmsmall]{acmart}
\usepackage{hyperref} 
\usepackage{hyperxmp} 
\usepackage{listings}
\usepackage{color}
\usepackage[frozencache=true,cachedir=minted-cache]{minted}

\definecolor{deepblue}{rgb}{0,0,0.5}
\definecolor{deepred}{rgb}{0.6,0,0}
\definecolor{deepgreen}{rgb}{0,0.5,0}

\usepackage[utf8]{inputenc}
\usepackage{CJKutf8}
\usepackage{makecell}
\usepackage{subcaption}
\expandafter\def\csname ver@subfig.sty\endcsname{}
\usepackage[most]{tcolorbox}
\usepackage{color, soul}
\usepackage{multirow, makecell}
\usepackage{tablefootnote}
\usepackage{multicol}
\usepackage{geometry}
\usepackage{minted}
\usepackage{makecell}
\usepackage[most]{tcolorbox}
\usepackage{color, soul}
\usepackage{multirow, makecell}
\usepackage{tablefootnote}
\usepackage{array, tabularx, caption, boldline}
\usepackage{graphicx}
\usepackage{colortbl}
\usepackage{color}
\definecolor{tabcolor}{rgb}{.126,.126,.126} 
\usepackage{float}
\usepackage{subfig}
\usepackage{soul}
\usepackage{booktabs}
\usepackage{caption}
\AtBeginDocument{%
  }

\setcopyright{acmlicensed}
\acmJournal{PACMHCI}
\acmYear{2025} \acmVolume{9} \acmNumber{2} \acmArticle{CSCW056} \acmMonth{4}\acmDOI{10.1145/3710954}

\newcommand{\RNum}[1]{\uppercase\expandafter{\romannumeral #1\relax}}

\begin{document}

\title{DanModCap: Designing a Danmaku Moderation Tool for Video-Sharing Platforms that Leverages Impact Captions with Large Language Models}

\renewcommand{\shorttitle}{DanModCap}

\author{Siying HU}
\affiliation{%
  \department{College of Materials Science and Engineering}
  \institution{Shenzhen University}
  \city{Shenzhen}
  \country{China}
}
\email{sying.ch1026@gmail.com}

\author{Huanchen Wang}
\affiliation{
  \institution{Southern University of Science and Technology}
  \city{Shenzhen}
  \country{China}
}
\affiliation{
  \department{Department of Computer Science}
  \institution{City University of Hong Kong}
  \city{Hong Kong}
  \country{China}
}
\email{hc.wang@my.cityu.edu.hk}

\author{Yu Zhang}
\affiliation{%
  \department{Department of Computer Science}
  \institution{City University of Hong Kong}
  \city{Hong Kong SAR}
  \country{China}
}
\email{yui.zhang@my.cityu.edu.hk}

\author{Piaohong Wang}
\affiliation{%
  \department{Department of Computer Science}
  \institution{City University of Hong Kong}
  \city{Hong Kong SAR}
  \country{China}
}
\email{piaohongwang@gmail.com}

\author{Zhicong Lu}
\affiliation{%
 \department{Department of Computer Science}
  \institution{George Mason University}
  \city{Fairfax}
  \country{United States}
}
\email{zlu6@gmu.edu}


\begin{abstract}
Online video platforms have gained increased popularity due to their ability to support information consumption and sharing and the diverse social interactions they afford. Danmaku, a real-time commentary feature that overlays user comments on a video, has been found to improve user engagement, however, the use of Danmaku can lead to toxic behaviors and inappropriate comments. To address these issues, we propose a proactive moderation approach inspired by Impact Captions, a visual technique used in East Asian variety shows. Impact Captions combine textual content and visual elements to construct emotional and cognitive resonance. Within the context of this work, Impact Captions were used to guide viewers towards positive Danmaku-related activities and elicit more pro-social behaviors. Leveraging Impact Captions, we developed DanModCap, an moderation tool that collected and analyzed Danmaku and used it as input to large generative language models to produce Impact Captions. Our evaluation of DanModCap demonstrated that Impact Captions reduced negative antagonistic emotions, increased users' desire to share positive content, and elicited self-control in Danmaku social action to fostering proactive community maintenance behaviors. Our approach highlights the benefits of using LLM-supported content moderation methods for proactive moderation in a large-scale live content contexts.
\end{abstract}



\received{January 2024}
\received[revised]{July 2024}
\received[accepted]{October 2024}

\begin{teaserfigure}
\begin{minipage}{\textwidth}
  \includegraphics[width=\linewidth]{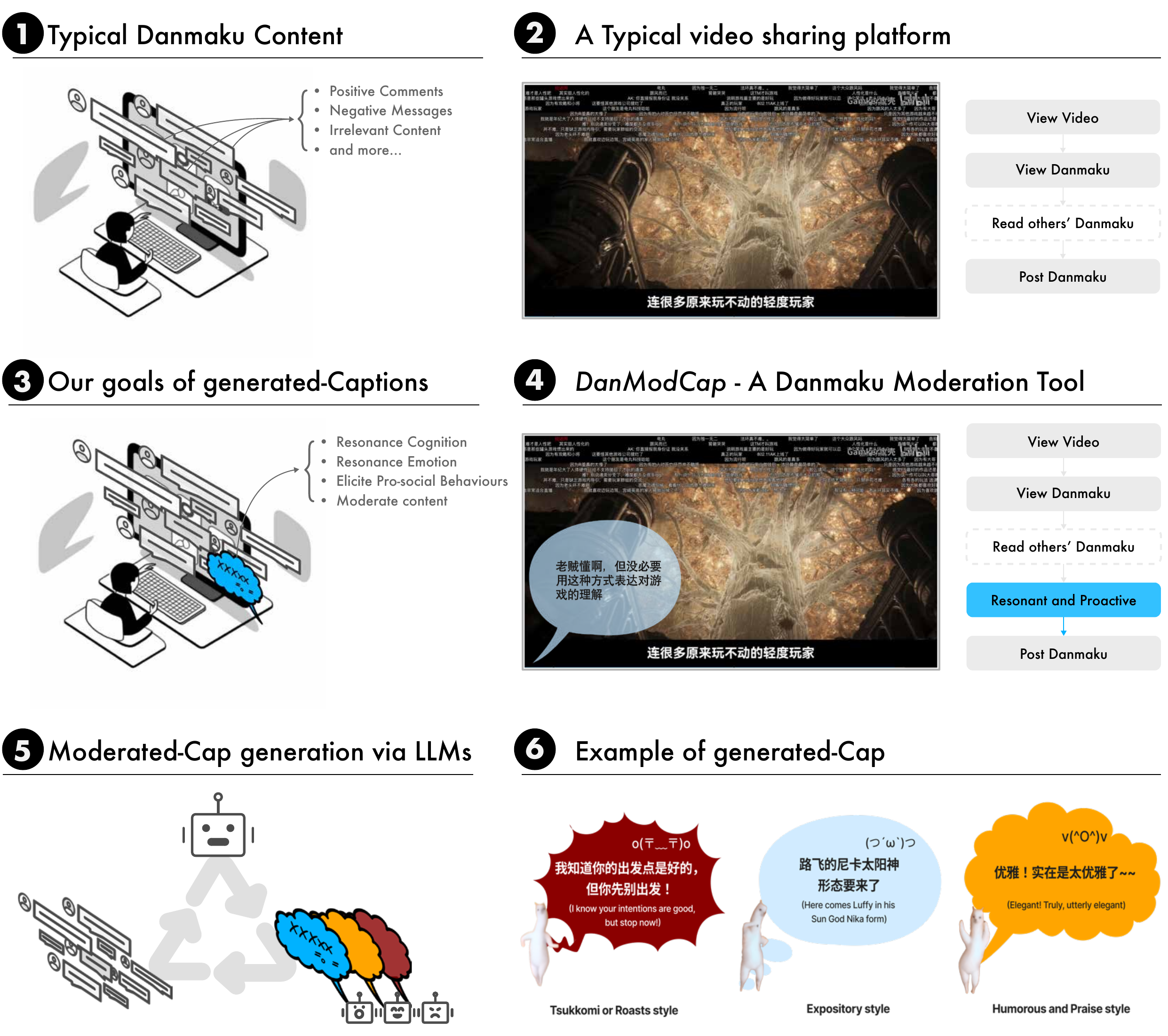}
    \caption{Our Danmaku moderation approach. 1) Typical Danmaku involves positive, inappropriate, or irrelevant content. 2) On video-sharing platforms, Danmaku is a built-in social feature that supports end-users' social communication and is synchronized with the video. 3) To moderate Danmaku content, we introduce a proactive approach called Impact Captions, that leverages a key feature in East Asian variety shows. 4) Building on this conceptual foundation, we introduce \textit{DanModCap}, a tool for Danmaku content moderation that uses generative AI to create Impact Captions. 5 and 6) The generation pipeline and example captions.}
  \label{fig:teaser}
  \end{minipage}
\end{teaserfigure}

\maketitle

\section{Introduction}
Online video-sharing platforms (VSPs) have become increasingly popular ways for individuals to consume and share information \cite{niu2023building}. To enhance user engagement, these platforms have several built-in social features. For example, users can provide comments or share messages within live chat rooms, or post \textit{Danmaku}. Danmaku is a real-time commentary method that overlays user comments containing internet slang and memes on videos to foster emotional expression and interaction among viewers \cite{wu2019danmaku,chen2022danmaku}. Danmaku's continuous presence in repeated video loops enhances the effect of each comment and shapes the viewer’s experience over repeated viewings. Its pseudo-synchronous nature is also a noteworthy departure from traditional forum-based communication, aiming to foster a sense of community while enhancing user participation \cite{wu2019danmaku,lin2018exploratory,li2017interface, zhang2020making}. This feature has thus become increasingly popular on popular video-sharing platforms such as AcFun, Bilibili, TikTok, and Twitch.

While Danmaku offers an immersive and socially interactive viewing experience, it also introduces challenges related to anonymity, the rapid spread of inappropriate messages  ~\cite{chen2017watching,bai2019stories,wu2019danmaku,he2021beyond}, comment overload, negative interactions, and language conflicts \cite{wu2019danmaku,zhang2020making}. There is thus a need for real-time Danmaku moderation. Moderating Danmaku, however, is a significantly different challenge than moderating non-real-time content. For example, the rapid and voluminous stream of Danmaku comments often leads to information overload \cite{wu2019danmaku} for users and moderators. The anonymity of expression also lends itself to increased user-to-user toxicity because the diversity and ambiguity of offensive or implicit harmful content complicates the moderation process. Existing moderation approaches also rely more on post hoc inspection, which makes them inappropriate for live settings \cite{niu2023building}. Moderation must also maintain a contextual understanding of Danmaku within a period of time so simply removing or blocking content without considering its context may distort the communication and interaction facilitated by Danmaku.

This research first conducted a video-content analysis to understand the existing uses of Impact Caption design. We identified a taxonomy of Impact Caption types that was organized into three categories and seven design dimensions. We then ran an expert-based co-design workshop to solicit design recommendations to steer Danmaku content toward the promotion of prosocial behaviors. Spurred by the expert recommendations and our taxonomy, we developed a Danmaku moderation tool, \textit{DanModCap}, that would support users in moderating live content while being mindful of community engagement and the dynamic nature of viewer pro-social interactions. A key tenant of \textit{DanModCap} is the \textit{Impact Caption}, a popular feature in East Asian variety shows \cite{basaran2013effects,sasamoto2014impact}. Impact Captions are visually compelling and context-sensitive textual and non-textual elements that are designed to captivate the reader's attention and enhance their emotional experience \cite{achlioptas2021artemis}. They are powerful triggers that add depth and meaning to content. Our implementation of Impact Captions was rooted in \textit{the resonance theory} ~\cite{decety2008emotion,mcdonnell2017theory}, which suggests that when we see or hear something that resonates with us emotionally and cognitively, it creates a stronger connection to us. By using familiar phrases to encourage cognitive resonance~\cite{kirsh2013embodied} and vivid colors and speech bubbles to encourage emotional resonance \cite{wan2024metamorpheus}, Impact Captions have the potential to bridge the gap between users and content, thus helping to shape more prosocial behaviors on VSPs (Video-Sharing Platforms). 

Within \textit{DanModCap}, the thematic information and emotional tendency of  Danmaku is analyzed via Latent Dirichlet Allocation and generative AI models including Llama2 create corresponding Impact Captions. Our system integrates contextual awareness, visual appeal, and cognitive and emotional resonance to influence viewer intentions, perceptions, interpretations, and the social practices of viewing Danmaku-embedded videos. \textit{DanModCap} not only filters out inappropriate content but also amplifies positive content to foster a sense of community and engagement.

To explore the influence of the generative Impact Captions intervention on viewers’ internal emotions, cognition, attitudes, and external social behaviors, we then conducted a user study. The study found that viewers experienced both cognitive and emotional resonance with the generative Impact Captions, perceiving them as aligned with their own thoughts, experiences, and intentions -- particularly around online activism and community belonging. The captions influenced viewers' attention dynamics, at times diverting focus from the main video content to the captions themselves. They also fostered a heightened sense of community cohesion, belongingness, companionship, and security, among other findings.
Due to these research activities, we thus contribute:
\begin{itemize}
\item A novel feedback method, Impact Captions, which can be used to moderate Danmaku comments and subtly shape viewer interpretations and emotions. 
\item The design and implementation of \textit{DanModCap}, a Danmaku content moderation system for video-sharing platforms that integrates Impact Captions.
\item A nuanced understanding of the role of cognitive and emotional resonance in video-sharing platforms content moderation. 
\end{itemize}

\section{Related Work}  
Of most relevance to our research is prior work on content moderation.
\subsection{Content Moderation on Social Media Platforms}
There is a considerable research gap in relation to the effective moderation of content that evades standard censorship measures. Whether this pertains to conventional social media or Danmaku on video platforms, the central objective of moderation remains consistent: to identify and neutralize detrimental content, ensuring user protection and a positive social environment. To date, research on content moderation has focused primarily on social media platforms, emphasizing the regulation of static textual content and dynamic multimedia elements. Indeed, existing studies have largely targeted the identification and mitigation of harmful, fraudulent, or hate speech within textual content  \cite{seering2017shaping,jhaver2018online,scheuerman2021framework} and elements of violence, pornography, or copyright infringement in multimedia materials \cite{gongane2022detection,galli2022regulation}. Within these studies, the moderation of interactive behaviors has been consistently identified as important in terms of identifying abusive behaviors \cite{jhaver2018online,chandrasekharan2017bag} or social engineering threats \cite{gongane2022detection,he2021beyond,scheuerman2021framework}. The primary objective of this field of study is to safeguard the content uploaded on social media platforms, ensuring that it complies with legal and ethical standards and discourages harmful behavior.

Despite recent advancements in algorithmic automation (see Section \ref{cma}), issues such as implicit offensive content and substituting inappropriate or sensitive information persist \cite{gorwa2020algorithmic}. When coupled with Danmaku’s real-time nature and its inherent anonymity and cultural specificity, moderation challenges when Danmaku is present exceed those seen in conventional social media. To address these additional issues, the present research shifts focus toward real-time VSPs.

\subsection{Content Moderation Approaches}
\label{cma}
Online content moderation is a critical aspect of maintaining safe and productive digital environments. Recent work has highlighted disparities in the impact of moderation, which has been found to frequently replicate offline social marginalization within online spaces \cite{klassen2023stoop,thach2022visible,freire2022understanding,schluger2022proactive}.

Previously, rule-based work has shown that incorporating various natural language processing techniques, such as sentiment analysis \cite{hettiachchi2019towards,pavlopoulos2017deeper} and text normalization, with machine learning methods \cite{rifat2024combating,son2023reliable} can enhance the effectiveness of automated moderation. These systems identify and filter inappropriate content using preset rules, such as using blacklists and regular expressions to detect and remove violations.

While traditional methods are simpler and less resource-intensive, they fall short in handling the complexity and volume of modern online content. To date, numerous automated content moderation technologies have been developed, including blocklist tools, text filters, and human-assisted screening techniques \cite{horak2021technological,cresci2018harnessing}. Nevertheless, each extant technology comes with its own limitations, many of which concern an inability to truly adapt to diverse user experiences and societal contexts \cite{law2009understanding}. As such, a move toward proactive moderation strategies has emerged, driven by the limitations of reactive methods. These proactive strategies, represented by designs such as CAPTCHAs \cite{seering2019designing}, Storychat \cite{yen2023storychat}, and the GLHF pledge \cite{brewer2020inclusion}, indicate a move toward subtle, persuasive approaches that encourage positive community participation while promoting inclusivity and equity.

Algorithms like Latent Dirichlet Allocation (LDA) have also been shown to provide more accurate and efficient content moderation \cite{voros2023web,kolla2024llm}. LDA is often used to extract hidden topic information from large amounts of text data as a topic modeling technology. Ammari et al., for example, transformed Reddit threads into a Bag of Words model and applied LDA to identify distinct topics that parents were likely to discuss anonymously \cite{ammari2019self} . Jhaver et al. utilized LDA to analyze the content of removal explanations on Reddit, aiming to understand how these communications impact user behavior \cite{jhaver2019does}. Inspired by this research, we used LDA to extract hidden thematic information from Danmaku comment text to identify the core themes of user discussions.

Other work has demonstrated the potential of LLMs for content moderation \cite{kolla2024llm,qiao2024scaling,kumar2024watch}. Kolla et al. introduced an LLM-based moderation tool, LLM-Mod, to automate the process of identifying and flagging content that violates community guidelines on Reddit \cite{kolla2024llm}. It used a multi-step prompting approach, summarizing and highlighting community rules, defining key terms within those rules, and generating hypothetical examples of rule-violating posts, for textual content moderation. Within \textit{DanModCap}, we harnessed a similar approach, along with prompt engineering, to produce modulated text and visual elements. Kumar et al. conducted an evaluation of the efficacy of LLMs in content moderation tasks, focusing on rule-based community moderation and toxic content detection \cite{kumar2024watch}. They found that while LLMs demonstrated comparable performance to human moderators in certain rule-based tasks, there was a marginal improvement in toxicity detection performance as model sizes increased, suggesting a potential plateau in the capability of LLMs for this moderation task.

Many of these automated methods, however, often fail to infer meaning from a sequence of sentences, and linguistic features like sarcasm, metaphors, and irony could result in user content being inaccurately flagged as inappropriate. Thus, although these methods are effective they struggle with the volume and complexity of modern online content \cite{nobata2016abusive}. To address the challenges posed by automated content moderation, another widely adopted approach is to integrate human intervention, such as community-based content moderation Mechanisms \cite{wang2024efficiency}, crowdsourcing \cite{hettiachchi2019towards} or employing professional mediators. This involves utilizing specialized human content moderation teams to execute the content moderation tasks. These implementations are more straightforward but less adaptable to the nuances of human language and context, which can be time-consuming and less scalable. Some inappropriate content may still be free of grammatical errors and sound very eloquent, making it difficult for even experienced moderators to spot at first glance.

\section{Study 1: Impact Caption Video Analysis}
\label{ImpactCaptionContent}
To better understand existing Impact Caption design, we first performed an in-depth video-based analysis of popular TV series featuring Impact Captions. Our analysis focused on the visual representation of Impact Captions, their contextual relevance, design considerations, and their role as guiding elements for increased viewer interpretation.

\subsection{Procedure}

We initially consulted three global internet film and television databases, i.e., IMDB\footnote{IMDB: \url{https://www.imdb.com/}}, TV Time\footnote{TV Time: \url{https://www.tvtime.com/}}, and Douban\footnote{Douban: \url{https://www.douban.com/}}, to identify shighly popular variety shows from East Asia that had an airing history of at least eight years and had consistently incorporated Impact Caption content. These included China’s \textit{Who is the Murderer}\footnote{Who is the Murderer: \url{https://en.wikipedia.org/wiki/Who\%27s\_the\_Murderer}}, South Korea’s \textit{Running Man}\footnote{Running Man: \url{https://en.wikipedia.org/wiki/Running\_Man\_(TV\_video)}}, and Japan’s \textit{VS ARASHI}\footnote{VS ARASHI: \url{https://zh.wikipedia.org/wiki/VS\%E5\%B5\%90}}. We then watched the top three episodes from each show that had the highest viewership to gain a preliminary understanding of the show’s content. While watching, we identified that Impact Captions had several diverse visual design features that influenced audience appeal and manipulation based on previously work \cite{sasamoto2014impact}. For example, we found that videos with Impact Caption employed a series of striking visual design elements, including vivid color contrasts, dynamic typography layouts, and theme-coordinated background images.

Following this, we examined YouTube\footnote{YouTube: \url{https://www.youtube.com/}} and Bilibili\footnote{Bilibili: \url{https://www.bilibili.com/}} to analyze current design trends and the utilization of Impact Captions in variety shows on these platforms as online video platforms allow for more experimentation and innovation in visual design compared to  more constrained television formats. Three researchers conducted keyword searches on these video-sharing platforms (e.g.,  ``variety show Impact Caption'', ``Impact Caption in the TV program'' and ``video Impact Caption'') to find and select representative samples based on the number of views. After filtering out irrelevant and duplicate content, 27 videos with an average duration of 43.23 minutes and 33.83k views were selected for analysis. The researchers then conducted independent open coding of a subset of the videos to identify critical aspects of Impact Captions, including visual elements, interaction patterns, and the perspective, using methods established in prior work (~\cite{sasamoto2014impact,lu2022postmodern}). Following multiple rounds of reflection and discussion, a taxonomy of Impact Captions was developed, with three types of captions that spanned seven design dimensions (\autoref{fig:ic_presentation}). The outcomes from this analysis were used to inform our subsequent research activities.

\subsection{Identified Caption Taxonomy}
Within our taxonomy, three categorizations of captions were identified.

\subsubsection{Caption Style 1: Visual Elements}
For the first categorization of caption, two primary categories were identified: Textual and Non-textual elements. Textual elements conveyed information like names and dialogue using typography, color, size, and orientation. Non-textual elements, used speech bubbles, emoticons, and ornaments to catch viewers’ attention and convey emotions. These categories collaboratively shaped Impact Captions’ visual and textual lexicon to engage viewers.

\begin{figure}[ht]{
    \includegraphics[width=\textwidth]{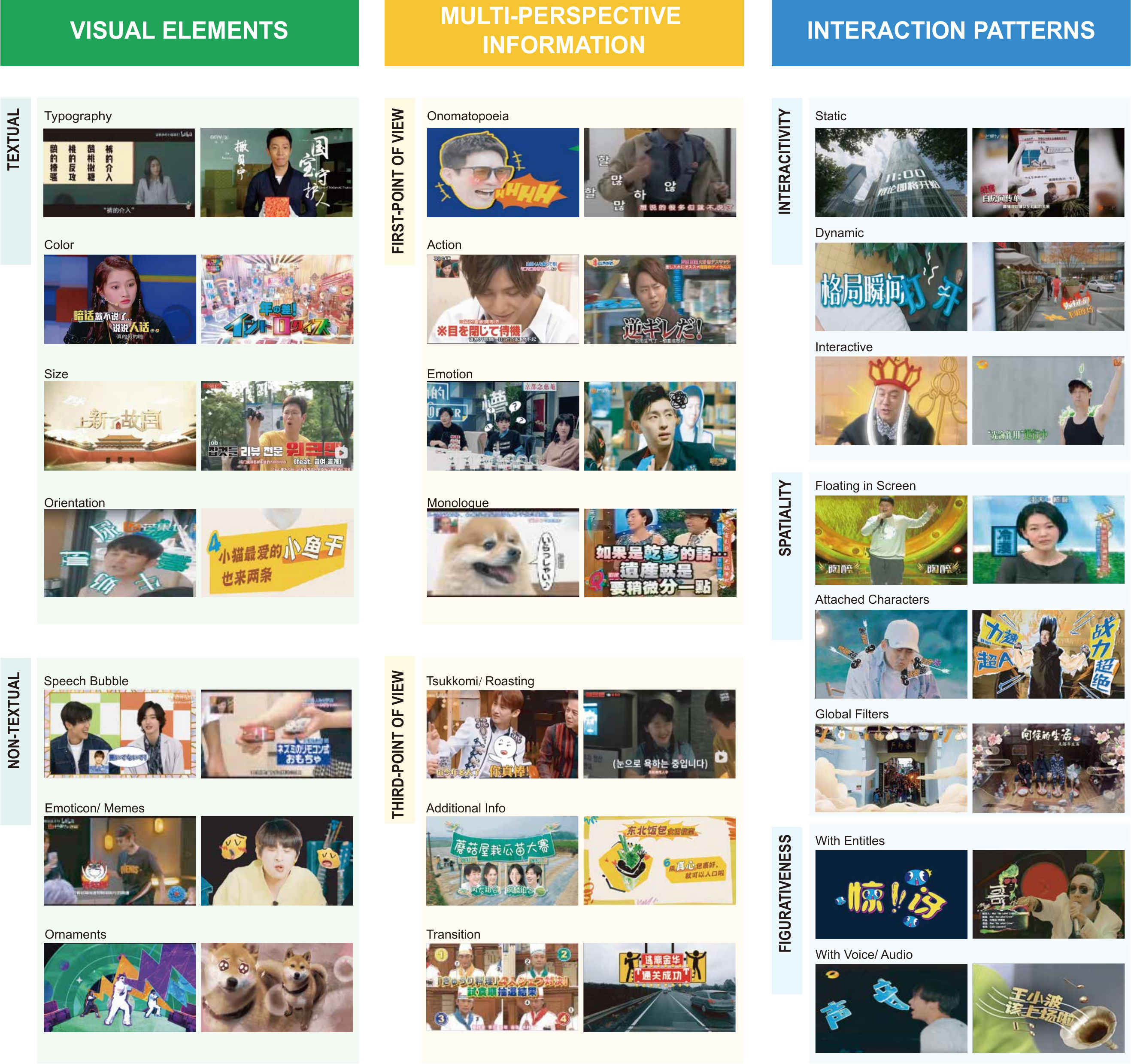}
    \caption{The Taxonomy of Impact Captions that was identified from our video-based analysis of popular online variety shows.}
    \label{fig:ic_presentation}
    }
\end{figure}

\subsubsection{Caption Style 2: Multi-Perspective Information}
With Multi-Perspective Information, two perspectives were used: first person points of view (POV) and third person POV. First person POVs centered on the protagonist’s perspective, directly disclosing their experiences, thoughts, actions, and emotions. These captions were instrumental in forging an empathetic connection between the lead character and the audience. In contrast, third person POVs offered a more comprehensive, ‘removed’ viewpoint, encompassing additional details, commentaries, and humorous response styles like Tsukkomi (Roasting), Expository or Humorous and Praise. This approach significantly broadened the entertainment value and deepened audience understanding, offering an enriched viewing experience.

\subsubsection{Caption Style 3: Interaction Patterns}
With Interactive Patterns, three methods facilitated video interaction: Interactivity, Spatiality, and Figurativeness. Interactivity was characterized by visual or animated elements within the video frame that created dynamic interactions. Spatiality adorned characters with wearable elements, integrating visual effects or animations that altered the spatial arrangement of information on the screen, including dynamic annotations following character movements. Figurativeness, used visual elements such as lyrics or sound effect descriptors in conjunction with auditory components to offer a multifaceted perspective, thereby enhancing the overall audiovisual experience of the viewer.

\section{Study 2: Expert Co-Design Workshop}
\label{ImpactCaptionDesign}
Once we developed our taxonomy, we conducted a co-design workshop to solicit design insights from professionals in video post-production that focused on how they would design Impact Captions to promote pro-socail behavior. These design insights were then used within our \textit{DanModCap} system to foster prosocial behaviors when using Danmaku. 

\subsection{Participants}
We utilized a purposive sampling approach to ensure that the individuals we recruited possessed the expertise and experience relevant to our research focus \cite{campbell2020purposive}. We had two inclusion criteria. First, all participants needed to have direct experience producing video impact captions as part of their video post-production work. Possessing such experience would enable participants to provide insights into the challenges involved in this specialized craft. Section, participants were required to have at least 1 year of experience in the video post-production industry. This ensured that they had accumulated rich practical knowledge and could provide in-depth perspectives on the challenges and needs in Impact Caption creation.

We recruited six participants for our co-design workshop. The group was comprised of three males and three females (all aged between 27 and 35) with backgrounds spanning broadcasting, television editing and directing, and visual design. Participants included motion graphic designers, video editors, and post-production supervisors, each of whom had hands-on experience incorporating video impact captions into their work. To avoid bias, our recruitment process intentionally omitted terms like ``variety show Impact Caption'' and ``Danmaku'', instead framing the work as ``Design and Research on Video Sharing Platforms''.  While the small number of participants could be considered a potential limitation, the depth of their industry expertise and specialized knowledge provided rich, context-specific insights that informed the subsequent stages of our research.

\subsection{Procedure}
We utilized a two-phase co-design session methodology that was employed in prior work~\cite{bartolome2023literature,xia2022millions,dym2022building}. Each phase lasted one hour.

\subsubsection{Phase 1. Group Interview for Sensitization}
The phase commenced with a researcher showcasing Impact Caption examples from popular shows to all participants to clarify the Impact Caption's design elements (\autoref{fig:ic_presentation}). Semi-structured interviews followed, focusing on each participant's personal experiences with video post-production, especially in Impact Caption design. This phase explored participant motivations and design considerations and provided illustrative examples of various Impact Caption applications.

\subsubsection{Phase 2. Designing Activity and Group Discussion}
After the interviews, participants worked together to develop Impact Captions that promoted positive social behavior when using Danmaku. They used the taxonomy derived in Section \ref{ImpactCaptionContent} to design Impact Captions that would create a healthy online environment. They used Danmaku videos and creative tools (such as Slides and Figma) to conceptualize their ideas and brainstormed ways to integrate Impact Captions into a Danmaku moderation tool. The conversation covered viewer engagement, perceptions, challenges, and potential solutions.

\subsection{Data Collection and Analysis}
With the participants’ consent, their engagement during the study was recorded and transcribed into text, and their design prototypes were collected. Then, using the methodology proposed by Chen and Zhang, open-ended coding of both the transcripts and prototypes from the workshops were undertaken \cite{popping2015analyzing,sasamoto2021hookability}. Two authors initially selected 20\% of the transcribed texts and prototype designs for open coding, reaching a consensus on the coding approach. Following this, two authors collaborated to code the remaining data.

The coding focused on components within each moderation design (e.g., challenges, Impact Caption elements, input, and output), workflows, contexts, and other visual elements (e.g., presentation style, textual content, and colors). This coding process drew upon existing research on visual elements, implications, and interaction patterns within video coding~\cite{chen2015remote}. Finally, one author categorized all coded components into themes.

\begin{table}[!ht]
\centering
\caption{The moderation and design challenges identified during the expert co-design sessions.}
\resizebox{\textwidth}{!}{
    \begin{tabular}{l|l}
    \toprule[1.2pt]
           \textbf{Challenge}& \textbf{Description}\\  
            \midrule[1.2pt]
            
    \multicolumn{2}{|l|}{\cellcolor[gray]{.8}Moderating Danmaku Content}\\   \midrule[0.8pt]
            Real-Time Nature and Scale & \makecell[l]{Respond to, and process, a large amount of real-time Danmaku information.}\\  \midrule[0.8pt]
            Complexity of Content & \makecell[l]{Understand and moderate Danmaku that contain implicit offensive content, slang, puns, and culturally-specific expressions.}\\ \midrule[0.8pt]
            Behavior& \makecell[l]{Viewers have different personalization preferences, which makes it difficult to handle their changing behavior.}\\  \midrule[0.8pt]
            Unpredictability of User Cognition & \makecell[l]{Viewers have different definitions of negative or irrelevant comments, which requires flexible moderation.}\\\midrule[0.8pt]
            Balancing Free Speech and Moderation & \makecell[l]{Finding a balance between the protection of free expression and appropriate regulation.}\\  \midrule[0.8pt]
         
           \multicolumn{2}{|l|}{\cellcolor[gray]{.8}Impact Caption Design}\\  \midrule[0.8pt]
           Content Relevance and Timeliness & \makecell[l]{There needs to be a close correlation and real-time updating of Danmaku content and the video's content.}\\  \midrule[0.8pt]
           Cultural Sensitivity and Appropriateness & \makecell[l]{Culturally insensitive or controversial content needs to be avoided.}\\ \midrule[0.8pt] 
           Visual Appeal and Readability Consistency & \makecell[l]{Impact Caption should be visually appealing while ensuring their message is easy to understand and consistent.}\\

    \bottomrule[1.2pt]
    \end{tabular} }
\label{summary_challenges}
\end{table}

\subsection{Results}
We identified several challenges that participants encountered while designing captions to promote prosocial behavior for Danmaku that related to moderating content and the design of captions themselves (\autoref{summary_challenges}). Moderation challenges included handling the scale and real-time nature of Danmaku comments, understanding nuances like slang and culture-specific expressions, managing unpredictable user behavior, and balancing free speech with moderation. Design challenges, meanwhile, focused on maintaining relevance and timeliness, avoiding culturally insensitive content, and ensuring visual appeal and consistency. 
\section{\textit{DanModCap} System Design}
\textit{DanModCap} integrates linguistic and visual strategies when generating Impact Captions to foster emotional resonance and encourage prosocial behavior. 

\subsection{Design Goals}
Based on the challenges identified during the co-design workshops (Section \ref{ImpactCaptionDesign}), we derived three design goals for \textit{DanModCap}:
\begin{itemize}

    \item {\textit{\textbf{D}esign \textbf{G}oal \textbf{1}:}} Support the proactive moderation of Danmaku through textual and non-textual Impact Caption elements to improve the efficiency of curation and reduce the need for post-video processing.
    \item {\textit{\textbf{D}esign \textbf{G}oal \textbf{2}:}} Encourage spontaneous prosocial user behavior for Danmaku-streaming content to reduce moderation burdens and improve the overall user experience.
    \item {\textit{\textbf{D}esign \textbf{G}oal \textbf{3}:}} Construct cognitive and emotional resonance via textual and non-textual elements while being mindful of relevance, cultural sensitivity, and visual appeal.
\end{itemize}

With \textbf{[DG1]}, we sought to actively manage Danmaku content on video platforms via several methods. First, by implementing an AI-based automatic content filtering system we wanted to analyze the content and emotional tendencies of the Danmaku in real time and identify and filter out undesired words, offensive speech, hateful speech, and inappropriate jokes. Second, the creation of a real-time monitoring mechanism, especially during peak hours or when popular programs are broadcast, would allow us to quickly identify and deal with problematic Danmaku information. Lastly, we wanted to understand community norms so we could convey clear Danmaku usage norms. 
 
With \textbf{[DG2]}, we wanted to guide users to actively abide by community norms by promoting positive and responsible social behaviors. We sought to realize this in several ways. First, through educational activities and guidance, we wanted to help users realize and understand the importance of community norms, encourage them to abide by these norms, and educate them on how to use Danmaku with responsibility and respect. Reward mechanisms (e.g., positive feedback or responses such as echoes and praises) were also designed to recognize users and comments that show positive prosocial behaviors. Users can be encouraged to participate in the process of Danmaku supervision by bringing personal perspectives to the role of community moderator and encouraging positive interactions between users to reduce negative or harmful Danmaku content.

With \textbf{[DG3]}, we will harness Impact Captions to nurture prosocial behavior within video platform communities. Our captions will be crafted to evoke emotional resonance through relatable text and graphics to encourage active community engagement, offering a subtle nudge towards thoughtful participation. Celebrating positive contributions through these captions will not only highlight model behavior but also set a communal standard, motivating users to contribute to a respectful and engaging dialogue. 

In summary, our approach is more than content policing as it aims to foster a community ethos where moderation is a collective endeavor, and positive engagement is the norm.

\subsection{Tenants of Impact Captions}
By leveraging verbal communication strategies and techniques to facilitate emotional support, Danmaku-inspired Impact Captions can function not only as a visual enhancement tool but also as a significant means of shaping Danmaku culture and fostering positive user behavior. Impact Captions can be thought of as resonant triggers that add an interactive narrative layer to content. They translate intangible elements such as emotional responses or social themes into tangible textual expressions to subtly guide viewer comprehension and emotional engagement.  
Impact Captions can also function as linguistic metaphors, playing a pivotal role in the classification and conveyance of emotions within Danmaku content. Such captions can be designed to align with the emotional states of viewers by employing commonly observed online communication styles. Within this work, we used the Tsukkomi (Roasts), Expository, and Humorous and Praise response styles. Each style serves a unique purpose when augmenting Danmaku as outlined next.

\begin{CJK*}{UTF8}{gbsn} 

\subsubsection{Generation Style 1:}
The \textbf{Tsukkomi (Roasts) style} is utilized with negatively oriented Danmaku content. It leverages an informal communication approach, commonly identifying loopholes or keywords in another’s language or behavior, followed by the expression of irony or doubt. According to interpersonal communication theory, this style employs humor to address critical comments while avoiding excessive negativity, thus preventing the treatment of viewers as ``emotional garbage'' \cite{hofmann2016interpersonal}. For instance, hyperbolic metaphors might be used to humorously address narcissistic tendencies, like ``这位自恋狂是不是以为自己是全宇宙的中心啊？你有皇位要继承吗? (\textit{Does this narcissist think they are the center of the universe? Do they have a throne to inherit?})''. This style facilitates the collective criticism of disruptive Danmaku content, promoting emotional regulation through shared complaints, which can engender a sense of unity and intimacy among strangers \cite{suls2002social,ellemers2002self}.

\subsubsection{Generation Style 2:}
The \textbf{Expository style} is targeted at Danmaku that are irrelevant or ambiguous and serves to redirect the viewer’s attention back to the video. The use of metaphors related to `adventures’ or `deciphering,’ such as ``前方高能！(High energy ahead.)'' or ``路飞的尼卡太阳神形态来了！(\textit{Here comes Luffy in his Sun God Nika form!})'', assists in clarifying the video content for the viewer.

\subsubsection{Generation Style 3:}
The \textbf{Humorous and Praise style} echoes and amplifies positive Danmaku content to foster a pleasant atmosphere and evoke positive emotions. It is useful in contexts that warrant enjoyable commentary, such as compliments. For example, in a vibrant travel vlog, employing phrases like ``神仙UP主 (\textit{God-tier UP})'' or ``优雅！实在是太优雅了！(\textit{Elegant! Truly, utterly elegant!})'' can create a lively Danmaku environment. This style enhances viewer engagement and strengthens the emotional connection between the viewer and the content, ultimately leading to positive viewer responses \cite{john2004healthy}.
\end{CJK*}

As highlighted in Section \ref{ImpactCaptionDesign}, employing both first-person and third-person perspectives in Impact Captions can also create a more intimate connection between the viewer and the video. The first-person perspective is instrumental in forging rich emotional connections as it can mirror personalized and emotional viewpoints akin to those a viewer might naturally express. This can cultivate an inclusive and cohesive atmosphere by aligning with the sentiments or inclinations presented in Danmaku comments. The deliberate choice of pronouns like "we" instead of "you" or "they" contributes to this inclusive environment, making viewers feel directly addressed and involved, thus enhancing their emotional engagement. The third-person perspective provides an objective lens, aiding viewers in comprehending the context of the video and the associated Danmaku comments. This perspective can employ analytical insights to generate comments that express objective and dialectical viewpoints. This approach can enrich the viewer’s understanding and experience by offering a broader perspective. It allows for a more detached and comprehensive interpretation of the content, facilitating balanced and informed engagement with the video. 

\subsection{Impact Caption Generation Model}
\textit{DanModCap} establishes a feedback loop by analyzing viewer reactions to Impact Captions and generating Impact Captions that adapt to changing viewer discussions and emotional dynamics. Within \textit{DanModCap}, there are three main steps for creating Impact Captions: (1) Identifying Danmaku Topics and Sentiment, (2) Analyzing Danmaku Context over Different Durations, and (3) Generating Impact Caption Text and Graphics.

\begin{figure*}
    \centering
    \includegraphics[width=1\linewidth]{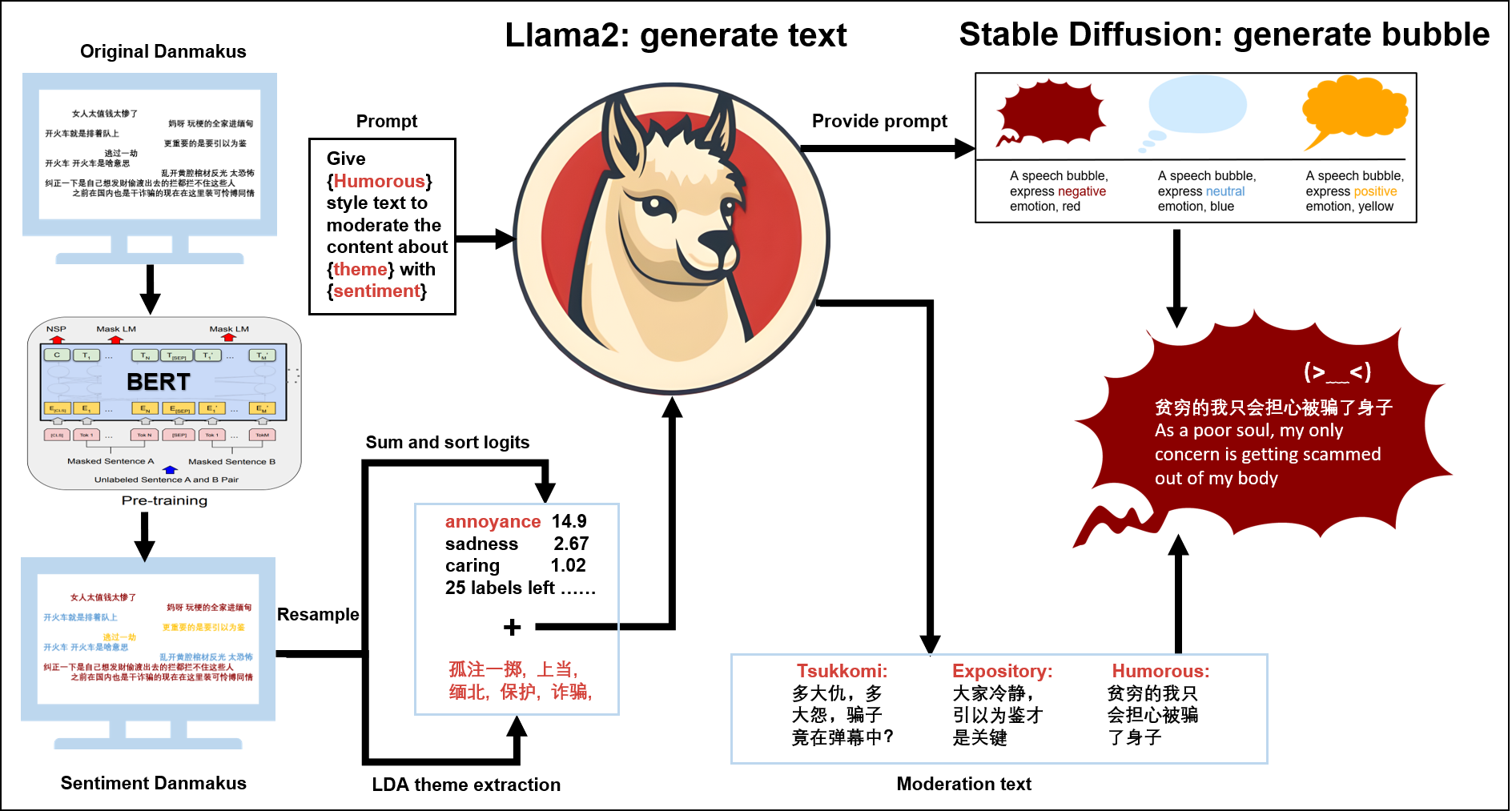}
    \caption{\textit{DanModCap}'s Impact Caption Generation Model, which used LDA to extract topics from Danmaku and supervised learning on a pre-trained model to identify Danmaku sentiment. Next, LLaMA2, which incorporates prompt engineering, was used to generate moderation text. Following this, Stable Diffusion generated a speech bubble.}
    \label{fig:pipeline}
\end{figure*}


\subsubsection{Step 1: Identifying Danmaku Topics and Sentiment}
To identify Danmaku behavior, the content and context of user discussions must be captured within a designated time interval and then the sentiment and expressed emotions must be evaluated to discern their behavioral patterns. Utilizing Latent Dirichlet Allocation (LDA) ~\cite{blei2003latent}, the main topics from each Danmaku were extracted. LDA was chosen due to its suitability to handle textual data, which is necessary for Danmaku comments. This method is particularly apt for scenarios involving large volumes of open-ended comment data, as it allows for the identification of the topics within the data without the need for predefined categories of themes.
Once the topics were identified, a multi-lingual fine-tuned BERT model ~\cite{pires2019multilingual}, trained on the GoEmotions dataset with 28 sentiment labels, was used to categorize the emotions expressed in the topics. This approach facilitated the comprehension of the content and sentiment of Danmaku discussions.

\subsubsection{Step 2: Analyzing Danmaku Context Over Different Durations}
Following the extraction of the content topics and sentiment, \textit{DanModCap} then assessed the overarching context shared by multiple Danmaku within a specified duration to create appropriate Impact Captions. Informed by insights from co-design session (Section \ref{ImpactCaptionDesign}), the duration for Impact Captions should be either 8 seconds or 12 seconds. As a result, separate resampling and aggregation techniques were applied to the Danmaku to construct the overall context and determine which duration was the most suitable. As each Danmaku can acquire 28 logits of sentiment labels (as per the multi-lingual BERT model), the summation of these values within the same duration enabled for the selection of the highest representative sentiment for that duration. An aggregation of the content from all Danmaku was then conducted and LDA was again employed to extract high-level semantic themes that encapsulated the context during that duration.

\subsubsection{Step 3: Generating Impact Caption Text and Graphics}
In Section \ref{ImpactCaptionDesign}, we found that speech-related visual representations were frequently used to visually convey information such as voice-overs or supplementary details. This finding led to the decision to adopt speech bubbles to elicit emotional connections from viewers, not just for aesthetics~\cite{aoki2022emoballoon,chen2021bubble} but also to foster a sense of community by enhancing expressiveness~\cite{lan2023affective,de2023visualization,stella2007bubble,chen2021bubble,aoki2022emoballoon}. 

In our design, speech bubbles have two essential components, a color and a shape. These elements should align with the emotional tone of the text to amplify emotional resonance and intensify viewers’ sense of empathy and connection. For speech bubble color, Expository-style Impact Captions employ a blue color to symbolize calmness while moderating irrelevant Danmaku. In contrast, the Humorous and Praise-style Impact Captions, which responds to positive Danmaku, utilize orange to evoke warmth and happiness. The Tsukkomi (Roasts)-style captions, which address more negatively charged Danmaku, use a dark red color to denote negativity. Translucent backgrounds are also used to maintain visual harmony with the video content.
For speech bubble shape, the shape boundaries have varying degrees of sharpness to reflect the nature of the conveyed emotions. For example, a lightning bolt-shape indicates that Danmaku is associated with negative emotions to communicate the intended emotional context.

To generate the Impact Caption text, we utilized LLaMA2 with prompt-tuning on the Chinese dataset ~\cite{pires2019multilingual} to generate text corresponding to the emotions and themes of Danmaku for the Chinese context \cite{kumar2024watch,nicholas2023toward}. We constructed the prompt to instruct the model to give the rational and regular output and used three response types (i.e., Tsukkomi, Expository, and Humorous/Praise) as input to evaluate the effect of moderation from the different response styles.

To generate the Impact Caption speech bubbles, we used LLaMA2 to generate prompts as input for Stable Diffusion (~\cite{ho2020denoising}) and used Stable Diffusion to generate visual speech bubbles. In addition, we mapped colors to sentiments for visual harmony. After generating the text and speech bubbles, we combined them with our script. The application of LLaMA2 not only enhanced the quality of the text generation but also improved the alignment between user emotions and video content, thereby increasing the accuracy and effectiveness of the generation.

By integrating LDA, LLaMA2, and Stable Diffusion, \textit{DanModCap} could understand and modulate the content of Impact Captions to handle the rapid flow of Danmaku and its cultural nuances. By using data pre-processing, topic extraction, and sentiment analysis (Step 1), alongside text generation (Step 3), and the intelligent design of visual elements (Step 3), \textit{DanModCap} should enable Impact Captions to resonate emotionally and visually with the audience.

\subsubsection{Assessment of Impact Caption Generation by Experts}
\label{ExpertAssessment}
To understand the efficacy of \textit{DanModCap}'s Impact Caption generation, we conducted a preliminary expert assessment with four professionals from video design, psychology, linguistics, and human-computer interaction. 

We first selected videos from the top five trending categories on the Bilibili platform, covering Games and Esports, Original, Anime, Car and Tech channels. From each category, we chose two videos that had at least 50,000 views and 8,000 Danmaku comments, and a duration of between 10 and 15 minutes. We initially collected 10 videos, each of which had a comprehensive range of response types (i.e., Tsukkomi / Roasts, Expository, and Humorous and Praise). To enhance the depiction of varied viewpoints and experiences, each video was classified into one of two categories based on its narrative perspective: first-person or third-person. This classification was necessary to generate two unique styles of Impact Captions for each video, guided by their respective Danmaku data. Consequently, the original set of 10 videos was expanded to 20 individual versions, each distinguished by its perspective and linguistic style, to enrich the diversity and depth of our experimental content. 
Impact Captions were then generated by \textit{DanModCap} for the videos. 

Each expert was required to view at least five videos with embedded Impact Captions, one from each video category. They then provided feedback on Impact Captions' emotional expression, visual design, and overall efficacy, adhering to established evaluation criteria. Interviews were conducted to gather experts’ opinions and recommendations to improve the Impact Captions. 

The opinions and recommendations underscored the model’s effectiveness in enhancing viewer engagement and providing emotional guidance. However, the timing and frequency of Impact Captions emerged as significant concerns. Experts noted that concurrently displaying multiple Impact Captions created visual clutter. They also highlighted how authenticity and accurately reflecting viewer sentiment was lost because popular Danmaku comments were not displayed alongside the Impact Captions. Lastly, to prevent implicitly harmful or offensive comments that may yield online conflicts, shifting from criticism to humor was recommended. 

The feedback showed that the language that was generated, particularly when responding to negative Danmaku with the Tsukkomi (Roasts) response style, needed to be revised. Thus, we revised \textit{DanModCap}'s generation to respond to negatively-valenced Danmaku with humor and jokes while reserving Tsukkomi-style language for positive content and certain ambiguous remarks. This adjustment sought to foster a more engaging and positive viewing experience.

\subsection{\textit{DanModCap}'s Interface and Implementation}
Following the refinement of \textit{DanModCap}'s Impact Caption generation model, we developed a web-based prototype platform that would enable moderators t0 modify categories, playback speed, and the screen placement of generated Impact Captions and viewers to customize Impact Caption types, sources, and visual designs based on their individual preferences.

\begin{figure}[htb]
    \includegraphics[width=\linewidth]{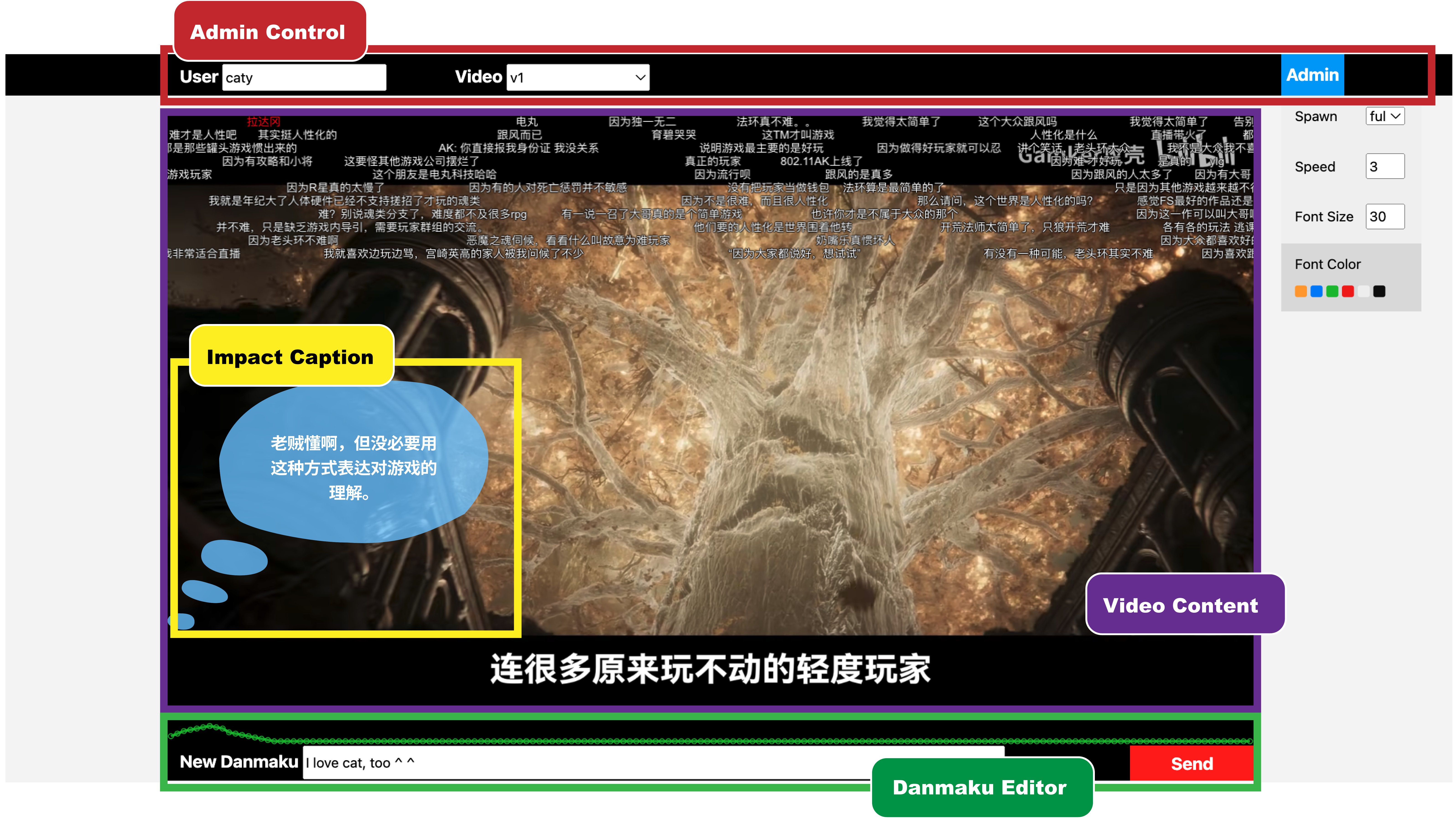}
    \caption{\textit{DanModCap} Interface including the Admin Control, an example Impact Caption, the Video Content, and the Danmaku Editor.}
    \label{fig:interface}
\end{figure}

The system’s interface was designed to replicate an authentic viewing experience, resembling Bilibili’s video playback page (\autoref{fig:interface}). By emulating Bilibili's widescreen mode, the interface minimizes potential distractions from secondary visual elements such as the sidebar and comments, thereby focusing on Danmaku comments and video interaction. The Danmaku Editor allows multiple viewers to transmit screen messages concurrently. These messages, displayed alongside the original Danmaku content, are essential input to the Impact Caption generation model. Using an adaptive visual pattern, Impact Caption's dynamically adjust their character size in response to the number of Danmaku messages within a specified time frame to alleviate the visual clutter mentioned during the expert assessment (Section \ref{ExpertAssessment}).

In the Admin Control, users can make real-time adjustments to the in-video display of Impact Captions. After logging in as a moderator, the user can adjust the Impact Caption settings ( including the Danmaku scale and comment threshold), choose various Impact Caption styles ranging from first person PoV Tsukkomi to third person PoV explanations (or a blend of the two), and define display positions, embedding methods, and Danmaku text obscuration preferences to customize the visual layout of Impact Captions.

\subsubsection{Implementation}
 Designed using native JavaScript, CSS, and HTML, \textit{DanModCap} used by Bootstrap~\cite{bootstrap2023} for its layout and Font Awe- some~\cite{fontawesome2023} for its icons. p5.js~\cite{p5js2023}  was also utilized to develop a layer to display the dynamic Danmaku and Impact Captions on the native video player. To interact with the Impact Caption generation model, the back-end used the Flask Framework~\cite{flask2023} with Python 3. Generated Impact Captions were stored as static files on a back-end server that also loaded various videos with preprepared Danmaku and performed basic tasks such as counting the number of real-time Danmaku.

\section{\textit{DanModCap} Evaluation}
\label{danmodcapuserstudy}
We conducted a user study to understand the influence of generative Impact Captions on viewers' emotions, cognition, attitudes, and social behavior. The study aimed to address the following research questions:
\begin{itemize}
    \item \textbf{RQ1:} Which emotions and information are conveyed by Impact Captions? 
    \item \textbf{RQ2:} How do users resonate with the emotions elicited by \textit{DanModCap}'s Impact Captions?  
    \item \textbf{RQ3:} How does \textit{DanModCap} impact users' perception of the Danmaku, the community, and Video-Sharing Platforms? 
    \item \textbf{RQ4:} Which social behaviors does \textit{DanModCap} elicit? 
    \item \textbf{RQ5:} Which practices do users follow when using \textit{DanModCap}? 
\end{itemize} 

\subsection{Participants}
We recruited 18 participants (i.e., nine male and nine female) from a local university to participate in our study. Snowball sampling was used for recruitment. Every participant reported watching and commenting on videos using Danmaku at least five times per week for over a year. All participants were between 20 to 29, aligning with the predominant viewer group (49\%) on Bilibili based on their published viewer statistics~\cite{bilibili2023}. Our participants were not compensated, and each experiment lasted approximately two hours.

\subsection{Study Stimuli and Apparatus}
Videos from five popular video genres were used in the study (i.e., Game \& Esports, Knowledge sharing, Lifestyle, Anime, and Short video) to increase the generalizability and reliability of the results and increase their appeal to different participants. We selected the top three most watched videos within each genre to use in the study. As we were going to create Impact Captions with two different POVs, this resulted in 30 videos being prepared (i.e., 5 genres x 3 videos x 2 PoV). To support real-time usage, \textit{DanModCap} was installed on the computers in a lab within a local university. It should be noted that as \textit{DanModCap} used pre-selected videos and only one participant participated at a time, participants were reacting to pre-generated Danmaku rather than Danmaku generated by other participants.

\subsection{Procedure and Task}
After providing informed consent, participants received training on how to use the \textit{DanModCap} system. They watched a 10-minute video on how to use the system. Then, they were tasked with contributing Danmaku comments to simulate the realistic usage of Danmaku on VSPs. Each participant viewed a minimum of five videos (at least one from each category), with 5-10 minute break in between each video. Video presentation was counterbalanced to prevent learning effects. 

After completing the video task, semi-structured interviews were conducted to understand the impact of generative Impact Captions on participants. Drawing inspiration from the Stimulus-Organism-Response Model embedded in the structure of Media Synchrony Theory~\cite{jacoby2002stimulus,li2019exploring,liu2016watching}, the interview questions between participants and the video content (\autoref{table:sormodel}). This theoretical framework allowed for a deeper exploration of the dynamics, interactions, and enhancing effects of generated Impact Captions on participants. Additionally, valuable information about moderation-related content design could obtained.

\subsection{Moderation During the Study}
During a pilot study with six participants, we found that participants had a difficult time posting spontaneous Danmaku comments while regulating their views and emotions, i.e., they aimed to maintain a neutral and unbiased tone in their comments to avoid revealing their personal values during the pilot study. Having an abundance of neutral Danmaku that do not require moderation unless they were irrelevant to the video content, reduced opportunities for \textit{DanModCap}'s features to be explored and experimented with. Addressing this required the integration of moderators into the experiment. Moderators' role was to artificially adjust the thresholds of emotional judgment and the generation of Impact Captions to deliberately favor more extreme content. This would then cultivate an atmosphere where the generation of Impact Captions intensified, leading to the formulation of actionable metrics to improve content moderation. The abundance of emotionally charged content would also act as a trigger, compelling participants to react and provide feedback, thereby prompting a wider spectrum of participant engagement and interaction to investigate the aforementioned research questions. This methodological choice also unveiled how various Danmaku types influence participants’ emotions, cognition, and behavior, thereby contributing to moderation practices.

\subsection{Data Collection and Analysis}
Following previous content moderation work ~\cite{yen2023storychat,jhaver2018online,ma2021advertiser}, we audio recorded each session and transcribed the recordings for later analysis. To gain insight into participants' perspectives, feelings, and behaviors, we performed a qualitative interpretive analysis, inspired by prior work \cite{jhaver2023personalizing}. 

\subsection{Results}
We begin by presenting findings relating to participants' perceptions of generative Impact Captions. We then focus on user understanding and engagement, extending beyond social and community activism to encompass user attention, community dynamics, and video platforms. We conclude with results related to moderation practices.

\subsubsection{Perceptions of Generative Impact Captions}
Herein, we explore participant's perceptions with respect to their perceived cognitive and emotional resonance and the role of textual elements in Impact Captions.

\paragraph{Perceived Cognitive Resonance} 
Our research findings suggest that participants’ cognitive resonance with Impact Captions stems primarily from their comprehension of the underlying intent in the captions’ creation. This comprehension was rooted in shared experiences and articulated intentions. A predominant segment of participants (i.e., P2-P11, and P13) reported a heightened understanding of Impact Captions and attributed this to a congruence between the content of the captions and their thoughts, attitudes, and intentions.  For instance, P9 remarked, \textit{``having followed LOL games extensively, I am intimately familiar with the discourse in Danmaku; when Impact Captions echoed sentiments about a player's subpar performance, it resonated deeply, reflecting a comprehension of the perspective of an e-sports aficionado.''} This alignment fostered a deeper connection and resonance with their experiences.

Furthermore, several participants noted how Impact Captions that mirrored their own opinions or values reinforced their connection to the content. P4 noted, \textit{`oObserving unanimous Danmaku praise in vloggers' videos and contributing similar comments creates a shared atmosphere. In these instances, I perceive the Impact Captions as reflecting a similar disposition and sense of belonging to a common community.''} Additionally, some highlighted how Impact Captions often articulated desires they themselves harbored but had not expressed, providing an alternative avenue for online expression, e.g.,  \textit{``I do not usually like to express myself online, but Impact Captions is kinda like a representative that can speak for me online''}(P1).

\paragraph{Perceived Emotional Resonance} 
Participants experienced emotional resonance when Impact Captions displayed emotions congruent with their own. Some participants (e.g., P4, P7, P9 and P13) often found themselves vicariously expressing emotions through these captions. For example, P9 shared discomfort when encountering derogatory online messages against a favored team, yet found solace in the Impact Captions’ empathetic response. As P9 expressed, \textit{``when I see messages online taunting and calling names to a team I support after they lose a game, it makes me uncomfortable, however, Impact Captions seemed to understand my emotions and responded to the others.''}

We also observed that some participants (i.e., P2, P7 and P15) were influenced by Impact Caption's visual elements. Emotions were often guided by the visual cues of the captions, such as the shape and color of speech bubbles in the Tsukkomi response style, which some participants associated with feelings of anger and repulsion towards provocative Danmaku. For example, P9 mentioned, \textit{``there were instances when I was furious at what the Danmaku were saying and was about to retaliate, but then the Impact Captions also appeared and expressed their disapproval, making me feel that they were just as disgusted as I was with the others' comments.''}

\paragraph{Interpreting Textual Elements in Impact Captions} 
All participants demonstrated an awareness of the purpose of text in Impact Captions and the context of their emergence. P3, for example, commented on the potency and appeal of the Tsukkomi response style finding it particularly impactful, ``\textit{This expression is so powerful and interesting, it is too funny. ''}
Other participants, like P1 and P8, perceived the tone of certain captions as mechanical and rigid or lacking distinctiveness, particularly in terms of the text. For example, P1 stated, \textit{``I understand the point of the text depictions, but I did not find anything special or different. Maybe the only difference is the length because these are all shown in the video. Maybe except for the different color and shape underneath, from a textual aspect, I did not find these Impact Captions to be different.''}

\subsubsection{User Understanding and Engagement}
Our study also uncovered several interesting findings related to participants’ attention, sense of community connection, and heightened awareness of VSPs' ability to utilize user data for content generation. It also shed light on how participants were cognizant of their social impact and responsibilities.

\begin{CJK*}
{UTF8}{gbsn}
\paragraph{The Dynamics of Attention} 
\label{attentionchange}
We identified a dynamic shift in participants' attention, where they were initially drawn to Danmaku, oftentimes at the expense of the primary video content. However, the introduction of Impact Captions gradually fostered a realignment of their focus either towards the captions or towards the video. This was particularly prominent with Expository response style captions, which discouraged overindulgence in Danmaku. As noted by P2, \textit{``I have actually forgotten what this video is for, you know, I think it is fine to listen to the voice-over of the video is all, the Danmaku is much more exciting, so I am mostly going to watch the content of the Danmaku and not pay attention to the video screen because the screen is not that important, I think. This Impact Caption popping up to tell me what the video is about is fine; if it is something I am interested in and I missed it, I will rewind and go back to check it out.''}  A subset of participants also reported that Impact Captions could divert their attention towards Danmaku content, e.g., \textit{``Initially I was unaware of what was going on with Danmaku, paying attention to see [the video creator] in make-up and explaining, and suddenly noticed what the Impact Captions were saying inside `注意言辞，尊重他人 (Pay attention to words and respect others)', and I paused to see what they were fighting about.''} (P4).

\paragraph{Sensemaking for Community Cohesion}
Our findings also suggested that participants experienced a heightened sense of community cohesion, characterized by feelings of belonging and security. We observed that certain linguistic constructs were used as a distinct and flexible language by some participants (e.g., P4, P9 and P13). This language played a crucial role in cultivating a sense of group belonging and shared interests. Impact Captions' capability to reflect participants' emotions and perceptions significantly contributed to reinforcing this sense of belonging. P13 remarked, \textit{``I have seen the King of Thieves video written to appear ``感谢凯多老师！ (Thank you, Kaido-sensei!).'' , `恭喜路飞新帝(Congratulations to Luffy Shindei!)' When I see the Danmaku video, I feel that it [Impact Captions] and I are both secondary characters who also watch anime and understand us and that everyone swipes such Danmaku.''} As P4 expressed, \textit{"the Impact Captions felt like a like-minded friend, accompanying and interacting with me throughout the video."}

Several participants (P1, P7 and P9) reported a sense of security when Impact Captions articulated their unspoken thoughts or emotions. P9 noted how Impact Captions provided a voice for her unexpressed thoughts: \textit{``there are times when I am afraid to post Danmaku for fear of being cyberbullied, especially by some fans, however, this [Impact Captions] makes me relax my stress, like an alternative to my expression, instead of expressing myself on a public platform, with the feeling of being protected.''}

\paragraph{Implications for Societal Ethos in VSPs Utilizing Generative AI}
We found some consensus on the societal responsibilities of video platforms, especially those employing generative AI technologies (i.e., P1, P2, P13 and P14) . Content on these platforms, ranging from entertainment to user-generated material, mirrors and potentially molds societal culture and values. Danmaku often serves as a forum for social discourse, influencing public consciousness and attitudes. Impact Captions can subtly guide participants' behavior by endorsing positive Danmaku interactions or critiquing negative ones. P1 reflected how, \textit{``the deployment of Impact Captions carries inherent social responsibility. These captions do more than convey information; they steer Danmaku trends. Positive captions can foster a constructive viewing atmosphere, whereas negative or sarcastic use might deteriorate it. Hence, platforms with such capabilities must exercise discretion in their generative use, considering the broader impact on the online environment.''}

\subsection{Moderation Practices}
We also identified several moderation practises that Impact Captions had an impact on (e.g., self reflection, self-regulation, sharing and debating desires, and the spontaneous upholding of a community's discourse).

\paragraph{Self-Reflection Facilitated By Danmaku Language}
Some participants (i.e., P7, P9 and P12) spontaneously self-reflected on the speech patterns and word choices they used in Danmaku. This introspection was primarily triggered by the perception that the generated Impact Captions magnified negative emotions and remarks, adversely affecting the viewing experience for others. For instance, P12 thought that the Impact Captions seemed to be responding directly to his Danmaku message, prompting a re-evaluation of his contributions. He reflected, \textit{``I would feel like that Impact Caption was there to troll me like another user saw my Danmaku and came over to remind me.''}

\paragraph{Emotional Self-Regulation and Resilience}
Other participants (i.e., P3, P4, P5, P7, and P14) reported that they deployed strategies for emotional self-regulation as a reaction to Impact Captions, particularly during heightened emotional fluctuation. While participants generally exhibited tolerance and restraint towards offensive or irrelevant Danmaku content, prolonged exposure to such content, and particularly that which involved personal insults, frequently provoked strong emotional responses. In these instances, Impact Captions served as a proxy for participants’ reactions, alleviating stress and facilitating an emotional equilibrium. For example, when participants encountered derogatory Danmaku, they perceived Impact Captions as a means of counter-expression, helping to alleviate their irritation and providing a sense of solidarity. 

In this context, Impact Captions were catalysts for emotional self-control. P7 illustrated this, \textit{``I am very uptight sometimes like that Danmaku kept saying that that blogger was unattractive, it is just a life attack. I get furious at that kind of thing, and the word `舔 (lick),' which girls especially hate, is very erotic, but a verb the system does not recognize as a sensitive word, so it cannot be blocked. Nevertheless, at this point, Impact Captions is like a way to help me fight against those Danmaku, or maybe I am not the only one who resents it. The mood was much better all of a sudden.''} Moreover, some participants noted that Impact Captions helped reorient their focus from Danmaku to video content, especially when encountering expository captions (as discussed in Section \ref{attentionchange}).

\paragraph{Transformation of the Sharing and Debating Desire}
Impact Captions were also found to influence participants' inclination to share and express opinions, fostering a more proactive approach to moderation. Reduced engagement in provocative, conflicting, or negative discussions characterized this shift. Some participants adapted their participation in Danmaku by reducing the frequency of their comments as Impact Captions echoed their sentiments, or transforming their originally negative or confrontational comments into positive contributions. P3 commented, \textit{``when Impact Captions articulate my thoughts, I no longer feel compelled to express them separately''}. Similarly, P8 and P13 thought that Impact Captions steered them toward more positive interactions, prompting them to delete negative messages and contribute more constructively. As P8 said: \textit{``I realized that the impact captions are trying to guide me, to make me do something else. To relieve my negative emotions, I will delete the negative message or something that people would not like I want to share and respond to the call of the Impact Captions, either going back to watching videos or just copying the roasting content said by the Impact Captions, it is really funny, to liven up the atmosphere.''}

\paragraph{Spontaneously Maintaining an Online Environment}
The affirmative nature of Impact Captions inspired several participants to actively support constructive Danmaku content. This included massive Danmaku that take over the screen, echoing positive comments, and encouraging engagement with video content. Some participants (i.e., P8, P11, P13, and P14) spontaneously contributed to fostering a healthy online discourse, particularly resonating with the positive community culture reflected in the Impact Captions. As P6 stated, \textit{``I made a conscious effort to maintain a positive tone in Danmaku, focusing discussions on video-related topics and employing neutral language to initiate new conversations.''}
While P11 explained, \textit{``I wanted to mimic the behavior of that Impact Captions, it made me feel like I wasn't alone in trying to be nice and talk normally online.''}

\end{CJK*}

\section{Discussion}
Our research uncovered several novel insights about Impact Caption usage, potential video-sharing moderation approaches, how to implement automatic moderation by leveraging emotional an cognitive resonance, and user opinions about such moderation techniques. Herein, we highlight our main findings as they relate to the challenges of Danmaku content moderation, the potential of Impact Captions, and the risks of automated content moderation tools.

\subsection{The Challenges of Danmaku Content Moderation on VSPs}
Danmaku’s real-time, succinct nature adds increased complexity to the moderation process, arguably exceeding the challenges encountered in traditional social media settings. Our strategy for handling Danmaku data included categorizing content based on its emotional sentiment  and responding within specific durations. Our experts noted the difficulties in processing large amounts of Danmaku and accurately addressing the diversity of comments that are made (Section \ref{ImpactCaptionDesign}). Similarly, participants in our \textit{DanModCap} user study (Section \ref{danmodcapuserstudy}) reported mixed experiences, with some expressing dissatisfaction with the system’s emotion recognition capabilities and others advocating for more nuanced engagement with content exhibiting negative
emotions. We also found that inappropriate or harmful Danmaku content can quickly proliferate, thus adversely affecting the viewing experience and the overall community atmosphere. 

Given these findings,  the successful moderation of Danmaku thus requires a deep understanding of the complex relationship between anonymity, freedom of expression, and the potential toxicity associated with anonymity. Implementing innovative moderation strategies that effectively address these challenges can go a long way in maintaining a healthy and constructive community on video-sharing platforms.

\subsection{Impact Captions as a Novel Moderation Approach}
This research introduced Impact Captions as a way to automatically and proactively moderate Danmaku content on video-sharing platforms. Impact Captions combine cognitive and emotional resonance principles with generative AI models to enhance the overall quality of social interactions when using VSPs. Although Impact Captions foster active user participation in moderation via bidirectional communication, it remains unclear whether this approach is effective at addressing all the challenges related to content moderation. 

Further, \textit{DanModCap} used generative AI models to craft ideogrammatic Impact Captions that were primarily in Chinese. While incorporating ideograms into Impact Captions can amplify cultural resonance and emotional expressiveness, our findings also indicated a need for more balanced and culturally sensitive approaches to content moderation on VSPs to effectively address the diverse cultural and communal perceptions of online harm. Devising a universally applicable content moderation policy, however, is a difficult task as it is compounded by diverse cultural, communal, and individual perceptions of online harm. 

\subsection{Risks with Automated Moderation Methods}
Our research has shown that applying affective computing can result in more emotionally resonant Impact Captions, however, when used for content moderation, it can also present several challenges that require continued research. One of the most significant challenges concerns the dual effects of emotional resonance. Although designed to align with viewers' emotions \cite{fang2018co,chen2019facilitating}, Impact Captions can inadvertently amplify negative sentiments, leading to extreme responses or emotional polarization. Therefore, it is crucial to balance the enhancement of viewer engagement with the risk of reinforcing negative emotional states.

Another challenge is the risk of creating content bubbles that are emotional or filter outside perspectives. Emotion-based Danmaku moderation can inadvertently surround users with uniform emotional content that may curtail their exposure to diverse viewpoints and emotions \cite{zhunis2022emotion}. Filter bubbles, which emerge from algorithmic content filtration, can lead users to only encounter information that aligns with their pre-existing interests and viewpoints \cite{chitra2020analyzing}.  It is possible that an over-reliance on algorithmic filtering in Danmaku moderation could exacerbate these bubbles, limiting informational diversity and potentially reinforcing user biases, thus obstructing open and varied communication.

Finally, an over-reliance on AI generated content poses a significant risk in relation to consistency and accuracy of output \cite{chen2023can,uchendu2023does}. The potential inaccuracies that can emerge while trying to generate complex human emotions and cultural subtleties, coupled with the reduced level of human creativity and authorship, can lead to undesirable user experiences.

\subsection{Summary}
The use of Impact Captions for content moderation was a novel approach to enhance user experiences on VSPs, however, it is crucial to evaluate the potential risks and challenges associated with emotional resonance, emotional and filter bubbles, and the use of LLMs. Future research leveraging Impact Captions and affective computing must optimize the benefits of affective computing while minimizing its adverse effects to ensure a balanced and responsible approach to content moderation.

\section{Limitations and Future Work}
Despite \textit{DanModCap}'s data-driven implementation, many data-driven technologies exhibit limitations when understanding user-generated text and its connection to other content. In the current implementation, we focused on using Danmaku text to generate Impact Captions, as this textual data provides valuable insights into viewers' emotional responses, attention, and cognitive states while viewing videos. However, our lack of consideration of the video content itself may result in Impact Captions that do not always accurately reflect all the contextual information available to viewers. This could potentially impact the consistency of the captions. To address these limitations, further research is needed to integrate video content analysis (such as analyzing video content using computer vision models) and multimodal sentiment analysis to enhance the contextual understanding and generated Impact Captions.

Within \textit{DanModCap}, whenever the weighted frequency of emotional comments surpassed a pre-determined threshold within a designated time interval, the system would initiate the generation of more extreme and diverse Impact Captions. Thus, the setting of this threshold played a pivotal role in determining how the system reacted to the intensity and nature of Danmaku interactions. Consequently, it had a significant impact on the overall dynamics of viewer engagement and content moderation. As such, the generalizability of our results should be carefully considered in light of this methodological choice.

Although \textit{DanModCap} was intended to convert viewer feedback and emotional empathy into visual feedback using emotional and cognitive resonance, its effectiveness depends on a viewer’s ability to express emotions and their cognitive understanding. Not all viewers may exhibit the same capacity to do so or willingness to share their emotions. Furthermore, because all participants were from the same local institution, there may be a potential for sampling bias, wherein participants' reactions to the generated Impact Captions may differ from those from other cultural backgrounds. Participants' familiarity and receptiveness to the concept of Danmaku, which has its origins in East Asian culture, may have also influenced their perceptions and suggestions regarding Impact Captions. This cultural predisposition could potentially limit the generalizability of our findings to more culturally diverse user groups who may have different viewing preferences and acceptance thresholds for video annotations.

While our \textit{DanModCap} evaluation (Section \ref{danmodcapuserstudy}) provided valuable insights into user perceptions and reactions to the \textit{DanModCap} moderation approach, the limited scale and scope of the current evaluation limits our ability to draw strong conclusions about the overall effectiveness of \textit{DanModCap}. The evaluation focused primarily on exploring user experiences and did not involve a larger-scale, quantitative assessment of \textit{DanModCap}'s outputs. Future work should conduct a more rigorous, large-scale evaluation that examines the deployment effectiveness of \textit{DanModCap} and the natural reactions of viewers via an in-the-wild methodology \cite{son2023reliable,falk2024moderation}.
\section{Conclusion}
This research introduced \textit{DanModCap}, an LLM-based tool that moderated Danmaku content on video-sharing platforms through AI-generated Impact Captions. \textit{DanModCap} aimed to achieve non-intrusive moderation through cognitive and emotional resonance. During a lab-based study, we found that Impact Captions positively impacted viewers’ emotions, cognition, attitudes, and external social behaviors. These findings shed light on the potential of automated, generative approaches to moderation on video-sharing platforms. Our approach leveraged the complex dynamics of online communities and provided new ways to build proactive content moderation through resonant experience design. 

\section{Acknowledgement}
\begin{CJK*}{UTF8}{gbsn} 
We are thankful to all reviewers for their constructive feedback, which significantly strengthened this paper. We also gratefully acknowledge the inspiration drawn from various variety shows (particularly ``Who is the Murderer 明星大侦探'') and video-sharing platforms, which sparked new ideas and perspectives for this research. In addition, we extend our appreciation to our lab-mates for their insightful discussions and support throughout the research and review process.
\end{CJK*}



\bibliographystyle{ACM-Reference-Format}
\bibliography{Main}

\clearpage
\appendix

\section{Appendix: Prompt Engineering--Constraining Output}
\begin{lstlisting}[language=Python,  basicstyle=\tiny]
def gen_text_template(emotion28, emotion_dict, topics, comm_style, video_title, time_interval):
    """Generating the moderation text by prompt template

    :param emotion28: fine-grained word (28 emotional types) to describe the emotion of the majority of danmaku
    :type emotion28: str
    :param emotion_dict: a dictionary representing commands related to the coarse-grained emotions (3 emotional types). Each dictionary has the following keys
        - 'emotion': a string representing the emotion of the danmaku
        - 'view': a string representing the moderation style based on the emotion of danmaku
        - 'action': a string representing the action of moderation based on the emotion of danmaku
    :type emotion_dict: dict
    :param topics: the main topics of the majority of danmaku
    :type topics: str
    :param comm_style: the communication style (3 types) to configure the generated text style
    :type comm_style: str
    :param video_title: the title of the video being moderated
    :type video_title: str
    :param time_interval: the time interval to configure the frequency of text-generating (moderation)
    :type time_interval: int
    :return: the generated moderation text
    :rtype: str
    """
    prompt = """"Assuming you are a moderator responsible for reviewing barrage content and regulating user behavior on a barrage video site. You notice that the barrages around {at_time} seconds of the current video mainly discuss {topics}, with an overall {emotion3} and a sentiment tendency towards {emotion28}. As a moderator, how would you generate an appropriate barrage in a {flower_type} language style, from the perspective of {view}, to {action}? Please output your suggestion as a string with a total word count of no more than 20 characters, ensuring only one output."""

    data = {
    "prompt": prompt.format(
        at_time=time_interval,
        emotion3=emotion_dict['emotion'],
        flower_type=comm_style,
        action=emotion_dict['action'],
        video_title=video_title,
        topics=topics,
        emotion28=emotion28,
        view=emotion_dict['view']
    ),
    "temperature": 0.2
    }
\end{lstlisting}
\clearpage

\section{Appendix: Example Questions Used In The DanModCap User Study}

\begin{table}[htbp]
\small
        \caption{Example questionnaire used during the semi-structured interview, which was based on the Stimulus-Organism-Response Model that is part of Media Synchrony Theory.}
    \centering
    \renewcommand{\arraystretch}{1.3}
    \begin{tabular}{c|p{0.4\linewidth}|p{0.3\linewidth}}
    \toprule
    \textbf{Section} & \textbf{Example Questions} & \textbf{Goals} \\
    \midrule
    \multirow{2}{*}{Stimulus} &  1.1 In the course of interacting with the interface, did elements (such as Danmaku comments or text-based and non-textual effects) influence or reshape the comprehension of the video content? & \multicolumn{1}{p{0.4\linewidth}}{The factors that impacted their perception of the video content.} \\
    \cline{2-3}
     & 1.2 After watching the video, what transformations did you notice in your understanding? & \multicolumn{1}{p{0.4\linewidth}}{The changes they noticed in their understanding.} \\
    \hline
     & 2.1 While watching the video and the generated Impact Caption, did you undergo particularly intense emotional responses (such as anger, happiness, or surprise)? Can you pinpoint the specific elements of the generated Impact Caption that triggered these emotional reactions? & \multicolumn{1}{p{0.4\linewidth}}{The viewer's cognitive, emotional, and physiological reactions and the factors that led to them.} \\
    \cline{2-3}
    \multirow{3}{*}{Organism} & 2.2 Considering your experience, did the generated Impact Caption contribute to an enhancement or hindrance in your connection with other viewers? & \multicolumn{1}{p{0.4\linewidth}}{How the Impact Caption influenced their social connections and interactions with others.} \\
    \cline{2-3}
     & 2.3 Did your emotional state impact your interpretation of the video content, and to what degree do you believe this influence occurred? & \multicolumn{1}{p{0.4\linewidth}}{The degree to which their video content interpretation was affected by their emotional state.} \\
    \hline
     & 3.1 To what extent did the visual content of the generated Impact Caption influence your attitude or perception of the related Danmaku or video content? & \multicolumn{1}{p{0.4\linewidth}}{How aesthetics influenced their perceptions.} \\
    \cline{2-3}
    \multirow{3}{*}{Response} & 3.2 To what extent did the generated Impact Caption alter your interpretation or emotional response to the Danmaku comments? & \multicolumn{1}{p{0.4\linewidth}}{How viewer responses were influenced by the Impact Caption.} \\
    \cline{2-3}
     & 3.3 To what extent did these changes motivate you to engage in actions, such as joining online communities or leaving Danmaku comments on the video? & \multicolumn{1}{p{0.4\linewidth}}{Which behavioral responses manifested.} \\
    \bottomrule
    \end{tabular}   

    \label{table:sormodel}
\end{table}

\end{document}